Controllable selective coupling of Dyakonov surface wave at liquid crystal based interface

Yan Li, Jingbo Sun[*], Yongzheng Wen, and Ji Zhou[*]

State Key Laboratory of New Ceramics and Fine Processing, School of Materials Science and Engineering, Tsinghua University, Beijing 100084, People's Republic of China

*jingbosun@mail.tsinghua.edu.cn
*zhouji@mail.tsinghua.edu.cn

**Abstract**
Highly directional and lossless surface wave has significant potential applications in the two-dimensional photonic circuits and devices. Here we experimentally demonstrate a selective Dyakonov surface wave coupling at the interface between a transparent polycarbonate material and nematic liquid crystal 5CB. By controlling the anisotropy of the nematic liquid crystal with an applied magnetic field, a single ray at a certain incident angle from a diverged incident beam can be selectively coupled into surface wave. The implementation of this property may lead to a new generation of on-chip integrated optics and two-dimensional photonic devices.

**Introduction**
Optical surface waves are special waves of the electromagnetic modes strongly confined at nanoscale, propagating at an interface between two different media [1-3]. Their featured unique physical behaviors, such as surface sensitivity [4-6], field localization [7-9] and tailored interaction with light [10-12], have greatly propelled the promising applications in sensing [13-15], near-field imaging [16-18], and subwavelength optics [19-21], especially the nano guiding which can shrink the conventional optical waveguides into two dimensional photonic circuits [22-25]. The initial way to accomplish the surface mode is to use surface plasmons excited in between metal and dielectric medium, which is very easy to obtain, however, always suffering the propagation losses and the strong dispersion due to the consisted metal [26-29]. In contrast, a different type of surface excitation is the so called Dyakonov surface wave (DSW) [30], which propagates at the interface in between two transparent media, at least one of which has to be anisotropic, thus making a lossless surface wave a realistic possibility. Although predicted three decades ago, it was first observed experimentally until 2009 [31] because of its particular requirement on the relation of the permittivities from these two consisted materials, which was not easy to achieve in practice with natural materials, but on the other hand, made the surface wave highly directional. To date, substantial efforts have been invested to achieve the DSW among different interfaces of uniaxial [32-33], biaxial materials [34-36] or even with the help of metamaterials which can supply a strong anisotropy [37-45], making its realization more and more promising, and then the steering of the DSW in a tunable way becomes a really desired characteristic but barely studied so far. Here, we introduce and demonstrate a unique behavior of DSW coupling at the nematic liquid crystal 5CB (NLC) / polycarbonate (PC) interface which exhibits a strongly selective input coupling controlled with magnetic field. The use of NLC can provide a much stronger anisotropy but also a tunability of the orientation that steers the surface wave coupling magnetically, which allows the potential for developing new types of reconfigurable or switchable devices in photonic circuits, that are otherwise difficult or impossible with the conventional technologies.

**Results**

**Theory**

A unique property of the DSW is its highly selective coupling along with high directional propagation mainly determined by the certain condition of the isotropic/anisotropic refraction indices at the two sides of the interface, which actually makes the DSW hard to observe, however, a distinguished way for wave manipulation. As shown in Fig. 1, an interface is formed in between an isotropic material ($n_c$) and a positive uniaxial material ($n_e > n_o$). With the condition of the materials indices: $n_e > n_c > n_o$, the interface can support DSW propagation with a wave vector of **K**.

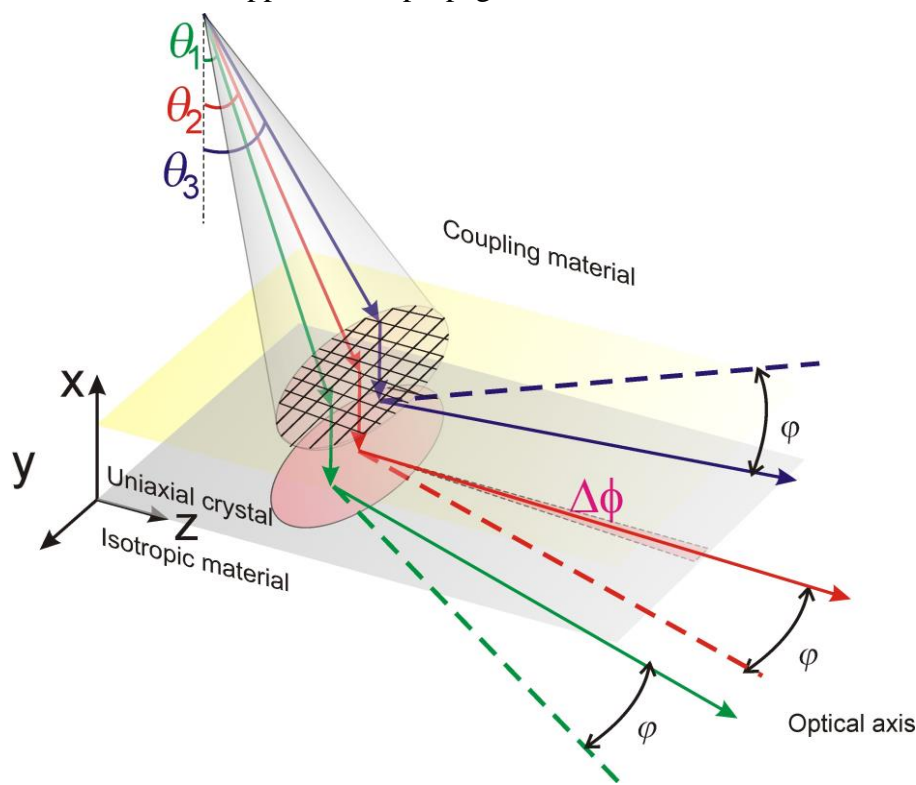

Fig. 1 Selective coupling behavior of DSW. A diverged beam is incident onto an interface formed by a uniaxial material an isotropic material. By meshing the illuminating region, the incident beam cone can be considered as a bunch of rays, each of which has its own identical wave vector and can be potentially coupling into DSW along the interface as long as the incident wave vector matches the wave vector of the DSW, which is only determined by the orientation of the optical axis. To coupling the three objective beams, optical axis is orientated to three different directions so that angle with respect to the wave vector is $\varphi$.

According to the Maxwell equation, the possible direction of **K** is determined with respect to the optical axis [30]:

$$\sin^2 \varphi_{min} = \frac{\xi}{2}\left\{1-\eta\xi+\left[(1-\eta\xi)^2+4\eta\right]^{1/2}\right\} \quad (1a)$$

$$\sin^2 \varphi_{max} = \frac{(1+\eta)^3 \xi}{(1+\eta)^2(1+\xi\eta)-\eta^2(1-\xi)^2} \quad (1b)$$

where $\eta = \frac{n_e^2}{n_o^2}-1$, $\xi = \frac{(n_c^2-n_o^2)}{(n_e^2-n_o^2)}$.

The modulus of $K=Nk_0$, expressed as a product of the effective index ($N$) and the wave vector in the air ($k_0$), can be determined by solving the transcendental equation [37]:

$$\tan^2 \varphi = \frac{\gamma_o(\gamma_c+\gamma_o)(n_o^2\gamma_c\gamma_e+n_c^2\gamma_o^2)}{n_o^2(\gamma_c+\gamma_e)(n_o^2\gamma_c+n_c^2\gamma_o)} \quad (2)$$

where $\gamma_o = \sqrt{N^2-n_o^2}$, $\gamma_s = \sqrt{N^2(\sin^2\varphi+\frac{n_e^2}{n_o^2}\cos^2\varphi)-n_e^2}$ and $\gamma_c = \sqrt{N^2-n_c^2}$.

Therefore, Eq. (1) and Eq. (2) define the direction and the modulus of the DSW's wave vector at the interface, both of which are determined by the orientation of the optical axis $\varphi$.

As shown in Fig. 1, to generate DSW, a s polarized beam with a certain divergence is incident onto the interface, which forms a beam cone. By meshing the illumination region, the incident beam can be considered as a bunch of rays from the source point to each joint of the mesh, such as the ray 1, 2, 3 in Fig. 1. Each ray has an identical wave vector which is uniquely determined by its incident angle. Inside the incident beam cone, only the ray matching **K** in both modulus and its direction can be selectively coupled into the interface as DSW, which helps us build a strict connection between the orientation angle $\varphi$ and the incident angle of the ray. Thus, by scanning the optical axis, we can select the certain ray from an incident beam as the input to couple into DSW and then transport it to the other objective direction at the output.

In order to manipulate the selective coupling, here we use NLC and a transparent PC material to form the tunable interface for the DSW. Under a bias magnetic field, the NLC exhibits a uniaxial positive birefringence as: $n_e=1.6889$, $n_o=1.5438$ with its axis aligned along the magnetic field [46]. Therefore, magnetic field can be used to control the orientation of the optical axis. A PC material with isotropic refractive index of $n_c=1.5788$ is chosen as the isotropic counterpart, and thus the interface satisfies the condition: $n_e > n_c > n_o$, which can be used to excite DSW [30].

**DSW demonstration**

The DSW coupling behavior is demonstrated based on the polarization-conversion reflection scheme (incident–reflected $s_{in}$–$p_{out}$) performed in a modified Otto-Kretchemann configuration consisted of ZnSe prism, NLC and PC resin block, where there are three interfaces from the top to the bottom: ZnSe/NLC, NLC/PC and PC/air, as shown in Fig. 2(a). The index of the ZnSe ($n_{prism}$=2.5891) is larger than $n_e$, and thus enables a total reflection when the internal incident angle is larger than the critical angle $\theta_c$ at the ZnSe/NLC interface, which creates an evanescent coupling into the NLC/PC interface. So in our setup, the incident beam is coupled from the anisotropic material to the interface of DSW. The prism/NLC/PC system was placed in between two sets of NdFeB magnet blocks offering 297mT magnetic field uniformly which was strong enough to fully align the NLC molecular [48]. According to Eq. 1, the allowed direction of **K** defined by the angle regarding to the optical axis has to be in the range from $\varphi_{min}$=30.84° to $\varphi_{max}$=31.35° ($\Delta\varphi$=0.51°). By solving Eq. 2, the modulus of **K** can be calculated and based on Snell's law, we can determine other key parameters, such as the effective index ($n_{eff}$) of the NLC at $\varphi$, critical angle $\theta_c$ and the internal incident angle ($\theta_{DSW}$) at which the wave vector of the incident ray can match **K**. All the values of these key parameters when $\varphi$ is varying in between $\varphi_{min}$~$\varphi_{max}$ are summarized in Table I.

TABLE I. Key parameters in DSW coupling system

| Name | Parameter | at $\varphi_{min}$ | at $\varphi_{max}$ |
|---|---|---|---|
| Orientation angle (°) | $\varphi$ | 30.84 | 31.35 |
| Effective index of NLC | $n_{eff}$ | 1.5783 | 1.5793 |
| Critical angle at Prism/NLC (°) | $\theta_c$ | 37.5602 | 37.5881 |
| Wave vector of DSW (**K**) | $Nk_0$ | $1.5788k_0$ | $1.5793\ k_0$ |
| DSW Coupling angle at Prism/NLC (°) | $\theta_{DSW}$ | 37.5741 | 37.5881 |

According to Table I, $\theta_{DSW}$ is just a little bit larger than $\theta_c$, which means the DSW coupling always occurs right after the total reflection. Therefore, in the experiment, the diverged incident beam is adjusted so that its central incident angle is around $\theta_c$ at the prism/NLC interface. With the divergence, the internal incident angle spreading from $\theta_{min}$ to $\theta_{max}$ cover both $\theta_c$ and $\theta_{DSW}$: $\theta_{min} < \theta_c \leq \theta_{DSW} < \theta_{max}$, which is split into 4 regions: $\theta_{min}$~$\theta_c$, $\theta_c$~$\theta_{DSW}$, $\theta_{DSW}$ and $\theta_{DSW}$ ~$\theta_{max}$, corresponding to reflection/refraction, total reflection, reflection/DSW and total reflection, respectively. Due to the anisotropy, either the refraction or the evanescent coupling to DSW is polarization sensitive and may cause a polarization change to the reflected beams. Part of the s polarized E field was transformed into p polarization [47], as shown in Fig. 2(a). However, the polarization states of the beams from total reflection regions stay the same as the incidence: s polarization. If we use a p-polarized analyzer to filter the s polarization (mainly the beams from total reflection regions) of the reflected beam, we will observe two bright spots, which is a typical image for the existence of DSW []. One is resulted from refraction region, which people called leaky mode [31], and the other one is due to the coupling of DSW. As shown in Fig. 2 b and c, along the z direction, the left one (z-) is due to the leaky mode and the right one (z+) is corresponding to the DSW. This two-spot pattern can be treated as a projection of the p polarization distribution at the the prism/NLC interface in yz plane. Therefore, we can demonstrate the selective coupling behavior by analyzing the movement of the DSW spot, where DSW coupling occurs, in the yz plane while the optical axis is rotating.

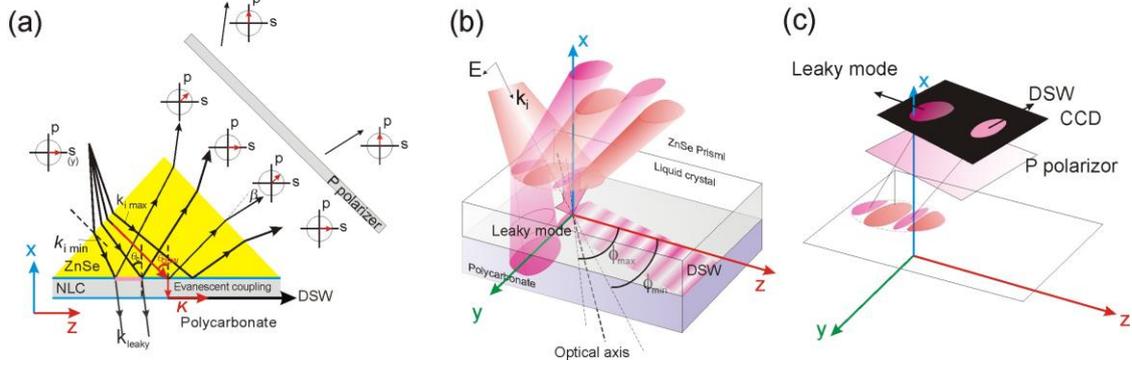

FIG. 2. Schematic of the DSW coupling with a diverged incident beam. (a) Since the divergence contains a series of wave vectors ($k_{i\,min} \sim k_{i\,max}$) including the leaky mode and the DSW coupling, the reflection is split into four beams and thus 4 regions: $\theta_c \sim \theta_{DSW}$, $\theta_{DSW}$ and $\theta_{DSW} \sim \theta_{max}$ with different polarization states. In this schematic, s polarization is the E field along y direction and p polarization means the E field is along the zx-plane. (b) In $\theta_{min} \sim \theta_c$, part of the beam transmits through the liquid crystal which is the leaky mode. At $\theta_{DSW}$, that part of the beam gets coupled into DSW at the NLC/PC interface. In the other two regions, there are total reflection. (c) According to polarization conversion theory, after the p-polarizer, we can see two bright spots: leaky mode (z-) and DSW (z+). By tracing this pattern back to the prism/NLC interface through the reflected beam, we can see the coupling region change with the rotation of the optical axis.

## Select coupling along z direction

In order to test select coupling along z direction, we first set the orientation angle of the optical axis to $\varphi$ with respect to z axis, as shown in Fig. 3(a). According to Table I, the incident angle was optimized to be 19.4° with a divergence of ±0.26° by a lens of $f$=150mm, which will finally result in incident angles from $\theta_{min}$=37.5336° to $\theta_{max}$=37.7247° along the zx plane at the Prism/NLC interface. Comparing to $\Delta\varphi$, divergence in y direction is small enough ($< \pm 0.1°$) that the y component of the incident wave vector ($k_y$) can be neglected. Therefore, we only need to consider momentum matching along z direction. We precisely control the rotation of the magnetic field from 30.7° to 31.4° and take a photo of the reflection beam after p analyzer by each increase of 0.1°. As shown in Fig.3b at the beginning ($\varphi$=30.7°), the reflection pattern shows one spot which means there is no DSW coupling. From 30.8°, two bright spots appear. According to our setup, the upper one is the DSW and slowly move towards z+ direction with the increase of $\varphi$. When $\varphi$=31.4°, one bright spot appears again which tells us the end of the coupling.

The movement of the pattern shows a selective coupling behavior along z+ direction, which can be interpreted as below. First, to ensure that rays in zx plane can be coupled, the optical axis is rotating in the range of $\Delta\varphi$ with respect to z axis. Second, by calculating Eq. 2, the increase of $\varphi$ results in the raise of $Nk_0$ in value, which requires a larger z component of the wave vector from the incident beam to match. Thus, rays of larger incident angle ($\theta_{DSW}$) in zx-plane are coupled into DSW in sequence, as shown in the inset in Fig. 3(b), which shows a upward movement of the two bright spot. When $\varphi$ is out of the range 30.8°~31.3°, the direction from the incident cone can not match $K$ and thus no DSW couple. The one spot is the leaky mode.

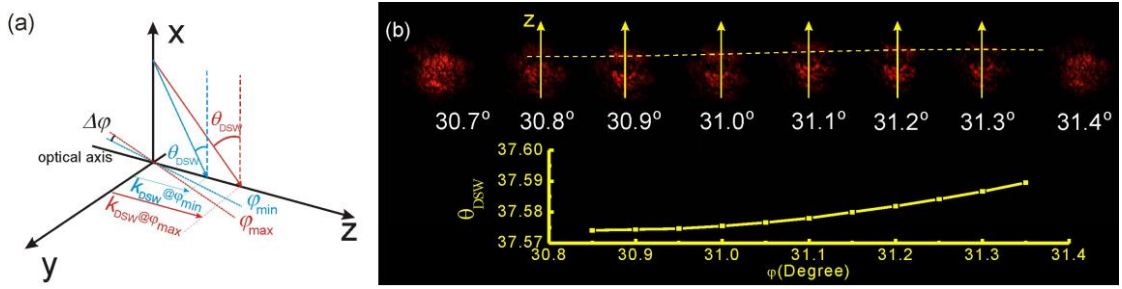

FIG. 3. Selective coupling behavior along z direction. (a) Schematic of the tunable DSW coupling. When the orientation angle is $\varphi_{min}$, the blue ray with small $\theta_{DSW}$ is coupled into DSW. When the oriating angle increases to $\varphi_{min}$, corresponding to a larger $Nk_0$, the red ray with a larger $\theta_{DSW}$ is coupled into DSW. (b) Observation of the tunable selective DSW coupling in zx-plane. Orientation angle is ranging from 30.7° to 31.4°. The yellow dash line indicates the upward movement of the spots in the camera. The ray with incident angle of $\theta_{DSW}$ can be coupled into DSW. The inset shows $\theta_{DSW}$ is raising with the increase of $\varphi$.

## Select coupling out of zx plane

Next, we test the selective coupling of the incident beam out of the zx-plane. As we mentioned, the beam's own divergence along y direction is too small to induce $k_y$ that can affect the momentum matching. Therefore, a small tilt angle $\phi$ is induced in between the incident beam and zx-plane artificially by adjusting a Mirror in front of the prism while $\varphi$ is set to be 31.2°. We carefully increase $\phi$ until the original two bright spot in Fig. 4(b) starts to merge into one from the left side of the pattern, as shown in Fig. 4(c). According to Fig. 4(a), due to this tilt angle, the wave vector has a component of $k_y$ which means the actual angle between new $k_{DSW}$ and optical axis is less than $\varphi_{min}$. Thus, there is no coupling of any rays from the incident cone. In order to couple this tilted beam into DSW, we further rotate the magnetic field so that the optical axis follows the titled beam until $\varphi$ falls into $\varphi_{min}$~$\varphi_{max}$ again. Finally, when we rotate the magnetic field to 33.2°, we can see that one bright spot starts to split into two bright spots again which means the angle of the optical axis is 33.2° with respect to the z as shown in Fig. 4(d).

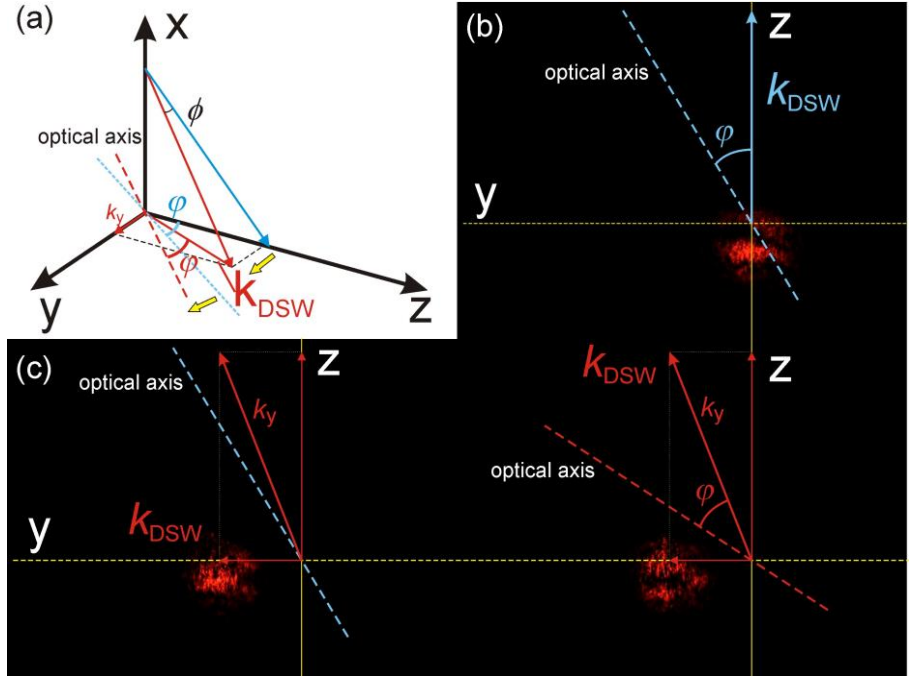

FIG. 4. Selective coupling of DSW out of zx-plane. (a) Schematic of the tunable DSW coupling when the incident beam is moving out of zx-plane by an angle of $\phi$. Lines in blue color indicate the original incidence in zx-plane and (b) the reflected pattern observed in the camera. The blue dash line labels the orientation of the optical axis. Lines in red color show the incidence with a tilt angle of $\phi$ and (c) the observed reflected pattern without rotating the magnetic field. The orientation of the optical axis is still the same with the one in (b), as labeled by the blue dash line. By rotating the magnetic field until the angle between the optical axis and the new wave vector is back to $\varphi$, the reflected pattern shows two bright spots again (d). The red dash line labels the new orientation of the optical axis. All the orientations of the axes are just schematic but not in actual values.

**Discussion:**

Based on the behavior above, we can make a scanning over each ray (with its distinguished wave vector) inside the diverged incident beam cone by rotating the optical axis and couple an expected one as the input to be the DSW.

In conclusion, we have experimentally demonstrated a selective coupling of DSW at the interface between transparent PC material and NLC. By studying the momentum matching condition at the interface, we build a strict connection between the orientation angle $\varphi$ and the incident angle of the ray. In the experiment, by rotating the optical axis of the NLC with an applied magnetic field, rays with different incident angles from a diverged beam can be selectively coupled into surface wave. The selective coupling property along with the highly directional lossless propagation of the DSW can be used to manipulate both the input and the output of a surface wave system simultaneously, which is readily integrated on a chip to form ultra-compact switchable devices.

**Materials and methods:**

Modified Otto-Kretchemann configuration

     The modified Otto-Kretchemann configuration is consisted of a ZnSe prism and a PC block with a layer of the NLC 5CB (Macklin P816665) in between. The thickness of the NLC layer is 5.36μm that determines the actual angle range of DSW which was directly measured by a optical microscope setup[49]. The anisotropic refractive indices were characterized by a total reflection method under the magnetic field of 297 mT (measured by CH-Hall 1600 Gauss/Tesla meter) along two orthogonal orientations. The measured $n_o$ and $n_e$ are equal to those measured from ellipsometer, which means the NLC is fully aligned under the 297mT magnetic field. Finally, the back surface of the PC block is grinded to form a rough surface to eliminate any possible back reflection from the PC/air interface.

**Acknowledgement**

This work was supported by the Basic Science Center Project of NSFC under grant NO. 51788104, as well as National Natural Science Foundation of China under Grant No. 51532004 and 11704216.